# Double Negative Metamaterials in Water Waves

Zixun Ge[1,#], Junke Liao[1,#], Linkang Han[1,#], Qilin Duan[1], Xiaofan Wang[1], Mengwei Dai[2], Shan Zhu[3,*] and Huanyang Chen[1,*]

1. Department of Physics, Xiamen University, Xiamen, 361005, China.
2. Institute of Electromagnetics and Acoustics, School of Electronic Science and Model Microelectronics College, Xiamen University, Xiamen, 361005 China
3. Department of Mechanical Engineering, Guangdong Technion-Israel Institute of Technology, Shantou 515063, China

[#]These authors contributed equally to this work.

Corresponding author. E-mail: *zhushanbay@gmail.com ; kenyon@xmu.edu.cn

## Abstract

Water waves present both opportunities and hazards, which demand precise control to effectively exploit their energy and mitigate their destructive effects. Leveraging the unique propagation characteristic of negative refraction enables versatile strategies for achieving such control. Here, we propose a Veselago-Pendry double negative metamaterial (DNM) for water waves constructed by nested gears and split tubes. This uniform array structure realizes effective negative water depth and gravity distributions, enabling tunable negative refraction that resolves the unclear structure-propagation relationships and stringent layout requirements of prior negative refraction structures. By employing coherent potential approximation (CPA), negative effective water depth $u_e$ and gravity $g_e$ are predicted. The predicted DNM parameters align well with band structures, and are validated by simulations of isolation, wave bending and all-angle imaging with surface waves excitation. A simplified experiment demonstrating water wave bending was successfully performed, matching the analytical predictions and simulation results well. Through quantitative mapping between structural parameters and propagation properties that enables tunable bandgaps and controllable negative refraction, DNMs furnish a transformative toolkit for coastal engineering, and are able to calm harbors, boost wave-energy harvesters, and steer river-bend currents to curb erosion.

## INTRODUCTION

Oceans and rivers hold vast amounts of wave energy but also pose significant threats to maritime safety, coastal infrastructure, and offshore operations. To address these challenges, various artificial structures have been developed to mitigate harmful wave energy or harness it efficiently. For example, wave isolation employs conventional breakwaters and permeable barriers to shield coasts(*1–7*), while resonance-based absorbers and unidirectional devices provide lighter, tunable protection (*8–13*). In contrast, focusing strategies amplify usable energy(*14*, *15*) through classic lenses(*16–22*) extended to innovative structures like cylinder arrays(*23*), and gradient-index lenses(*24*, *25*), enhancing wave localization and conversion

efficiency(*26–28*). These advanced functionalities arise from structures precisely engineered to control water wave propagation, which makes tailoring such designs to specific propagation behaviors essential for broadening their practical applications. Among wave propagation behaviors, negative refraction stands out due to its reverse deflection effect, which not only enables imaging and energy concentration(*29–32*)，but also facilitates wave isolation through wave steering(*31, 33–35*), greatly enhancing the flexibility of water wave manipulation. Early negative refraction of water waves was achieved primarily through photonic crystals, most notably vertical cylinder arrays(*29, 30, 36*) and checkerboard lattices of nearly touching square piles(*37, 38*) . Negative refraction in photonic crystals relies on wavelength-scale periodicity, which is impractical to deploy and lacks well-defined effective parameter. These limitations have shifted researchers' attention to metamaterials, which are composed of subwavelength structures and have opened new avenues for wave control(*39–42*). Negative refraction in water waves has since been realized with metamaterials such as channel networks(*43*) and surface-embedded tilted-plate grating(*31, 44*). More recently, strongly anisotropic hyperbolic metamaterials have provided a new approach for controlling negative refraction in water waves.(*33–35*). While existing negative refraction structures enable valuable water wave control, their practical deployment faces challenges arising from strict structure requirements, and low structural symmetry. Moreover, the relationship between structure and propagation properties remains incompletely defined, which hinders adjustable control in practical application. This motivates the pursuit of an alternative negative refraction strategy via water-wave double negative metamaterials, characterized by simultaneous negative effective depth and gravity, to enable precise tuning and robust control.

The concept of double negative metamaterials was first proposed by Veselago, yet no such material exists in nature. After Pendry *et al.* pioneered the realizations of negative permittivity $\epsilon$ via periodic metal wires(*45*) and negative permeability $\mu$ through split-ring resonators(*46*), Smith *et al.* experimentally proving the feasibility of double negative metamaterials(*47*) by combining them together. Double negative metamaterials with negative permittivity and permeability exhibit opposite phase and group velocities(*48*), yielding negative refraction(*49–51*), reversed Doppler effect(*52–55*), and inverse Cherenkov radiation(*56, 57*) for electromagnetic waves. Similarly, after achieving negative mass density $\rho$ with rubber-coated lead spheres(*58*) and a negative bulk modulus $\kappa$ with Helmholtz resonators in acoustics(*59*), researchers built the first acoustic double negative metamaterial(*60*), which displayed negative refraction(*60*), hyperbolic dispersion(*61*), and reversed Doppler effect (*62*) for acoustic waves. In water waves, the concept of negative effective gravity was proposed as early as 2011 through periodic resonant split tube arrays (*63*), and the realization of negative effective depth only came in 2025 with periodic gear arrays(*64*). Despite separate structures achieving negative gravity or negative water depth, the key challenge remains in integrating them into a novel design that balances their structural parameters to make both effective parameters negative simultaneously. Additionally, it requires precise quantitative mapping between structural parameters and the effective propagation properties.

In this Paper, we design a gear-in-split-tube structure that yields a double negative metamaterial (DNM) in water waves. Based on coherent potential approximation (CPA) (*65, 66*), the effective gravity $g_e$ and the reduced effective water depth $u_e$ can be obtained via a novel Hankel function approximation, enabling accurate scattering analysis while avoiding multivalued complex solutions. The predicted effective parameters agree well with the propagation characteristics inferred from DNM band structure. Full-wave simulations validate the DNM's propagation properties by demonstrating tunable negative refraction with wave bending and all-angle imaging in the double negative frequency range, as well as water wave isolation in the single-negative range. The surface waves excited by imaging further confirm the double negative properties and enable subwavelength resolution. Anomalous wave bending was experimentally observed via a simplified DNM array, which agree well with the analytical predictions and simulation results. By tuning geometrical parameters, DNM allows for selective control of wave isolation, negative refraction, and normal refraction of water waves at a specific frequency. Based on quantitative mapping between structural parameters and propagation properties, DNM can leverage tunable water wave manipulation like negative refraction and isolation, which offers high adaptability and inverse design flexibility for coastal engineering applications such as harbor calming, wave-energy harvesting enhancement, and river-bend current steering to mitigate erosion.

**RESULTS**
**Theoretical analysis and numerical simulations**

The double negative feature of the proposed structure is demonstrated through negative refraction, as shown in the schematic of Fig. 1(A). A periodic DNM array exhibits reverse wave propagation, leading to counterintuitive perpendicular wave emergence under a triangular arrangement. The DNM unit consists of two components, and the structural detail can be observed in Fig. 1(B) for the top view and Fig. 1(C) for the side view. The inner component is a gear-shaped cylinder with a central rigid core of radius $R_1$ with outer gear ring of radius $R_2$. The gear has $N_i$ notches, each with an arc length of $b = d\gamma$, where $d = 2\pi R_2/N_i$ is the arc length of one segment, and $\gamma$ denotes the ratio of the opening within a single cycle. The background water depth is $h_0$, while the water depth inside the notches is $h_1$. The outer component comprises split rigid tubes with inner radius $R_3$ and outer radius $R_4$. Each tube contains $N_o$ evenly distributed splits, with individual width $\Delta/N_o$, where $\Delta$ denotes the total split width. The dispersion relation of water waves(*67*) is

$$\omega^2 = gk\tanh(kh), \qquad (1)$$

where $\omega$ is the angular frequency ($\omega = 2\pi f$, $f$ is the frequency), $k$ is the wave vector ($k = 2\pi/\lambda$, $\lambda$ is the wavelength), g is the gravity acceleration, and $h$ is the water depth. Let the static water surface lies at $z = 0$, the vertical displacement $\eta$ of the water surface is obtained by $\eta = \frac{i}{\omega}\frac{\partial}{\partial z}\Phi(x, y, z = 0)e^{-i\omega t}$, where $\Phi(x, y, z)e^{-i\omega t}$ is the velocity potential. The *x*-component of the horizontal velocity $V$ is related to the displacement $\eta$ by $\partial V/\partial t = -u\, \partial \eta/\partial x$, where $u$ is the reduced water depth, defined as $u = [\tanh(kh)]/k$(63). Assuming the fluid is inviscid, incompressible and irrotational, the propagation of water waves is determined by the 3D Laplace equation(2, 68):

$$\nabla^2 \Phi = 0, \tag{2}$$

with the boundary conditions:

$$\begin{cases} \dfrac{\partial}{\partial z} \Phi = \dfrac{\omega^2}{g} \Phi \text{ on } z = 0 \\ \dfrac{\partial}{\partial \boldsymbol{n}} \Phi = 0 \text{ on surface I} \end{cases}. \tag{3}$$

Surface I represents the interfaces between the water and the fixed rigid body, with a unit normal vector $\boldsymbol{n}$. When vertical displacement $\eta$ is much smaller than wavelength $\lambda$ and water depth $h$, the Eqs. (1-3) will be simplified to a 2D equation

$$\nabla \cdot (u \nabla p) + \frac{\omega^2}{g} p = 0, \tag{4}$$

where $p$ is the the hydrostatic pressure, defined as $p = \rho g \eta$, with $\rho$ being the fluid density. When the wavelength $\lambda$ is larger than around four times of the periodic length $a$ ($\lambda > 4a$), DNM can be approximated as a homogeneous fluid with an effective gravity $g_e$, and effective water depth $h_e$ via CPA(65, 66), which results in dispersion relationship (1) being:

$$\omega^2 = g_e k_e \tanh(k_e h_e), \tag{5}$$

where $k_e$ is the effective wave number. The wave equation is

$$\nabla \cdot (u_e \nabla p) + \frac{\omega^2}{g_e} p = 0. \tag{6}$$

With the aid of the CPA, the effective parameters of DNM can be derived. The analysis focuses on the individual unit cell, where the surrounding DNM beyond radius R is treated as an effective homogeneous fluid with parameters $g_e$ and $h_e$. The radius $R = a/\sqrt{\pi}$, where $a$ is the lattice constant of the periodic arrays (shown in Supplementary materials Fig. S1), and the corresponding filling fraction is given by $\pi R_4^2/(\pi R^2)$.

Based on the structural symmetry, the vertical displacement $\eta$ of the water wave in cylindrical coordinates $(r, \phi)$ can be expressed by cylindrical wave expansion, and detailed analysis is shown in Supplementary material Section One. Notably, the subwavelength scale of the gear notches allows only the fundamental radial waveguide mode to be retained in the region of $(R_2 \leq r \leq R_1)$(64, 69). When the wavelength is much longer than the split width ($k\Delta/4 \ll 1$), the rigid tube can be replaced by an effective two-layer shell(42, 63). The inner layer ($R_3 < r < R'$) is an effective liquid with reduced water depth $u_3 = u_0/n_{st}^2$ and thickness $(R_4 - R_3)/n_{st}$, where $n_{st} = 2\pi R_4/\Delta$. The wave number in this region is $k_3 = n_{st} k_0$, while the gravity remains unchanged. The boundary is $R' = R_3 + (R_4 - R_3)/n_{st}$. The outer layer matches the fields of the interior ($r = R'$) and exterior ($r = R_4$) interfaces, maintaining the consistency of the total wall thickness (shown in Supplementary Materials Fig. S1(B)). As there is no scattering waves in region ($r \geq R$), $\eta$ can be derived as:

$$\begin{aligned}
\eta_1 &= [M_1 H_0^1(k_1 r) + M_2 H_0^2(k_1 r)] & (R_2 \leq r \leq R_1) \\
\eta_2 &= \sum_m [K_m J_m(k_0 r) + L_m H_m^1(k_0 r)] e^{im\theta} & (R_3 \leq r \leq R_2) \\
\eta_3 &= \sum_m [G_m J_m(k_3 r) + I_m H_m^1(k_3 r)] e^{im\theta} & (R' \leq r \leq R_3), \\
\eta_4 &= \sum_m [E_m J_m(k_0 r) + F_m H_m^1(k_0 r)] e^{im\theta} & (R \leq r \leq R_4) \\
\eta_5 &= \sum_m [A_m J_m(k_e r) + 0] e^{im\theta} & (R \leq r)
\end{aligned} \quad (7)$$

where the $m$-order Bessel function $J_m$ and the second kind of zeroth-order Hankel function $H_0^2$ describe inward propagation modes, and the first kind of $m$-order Hankel function $H_m^1$ represents outward propagation modes. The boundary conditions for $\eta$ are given as follows:

$$\begin{cases} p_i(R_C) = p_j(R_C) \\ \frac{iu_i}{\omega\rho}\frac{\partial p_i(r)}{\partial r}\bigg|_{r=R_C} = \frac{iu_j}{\omega\rho}\frac{\partial p_j(r)}{\partial r}\bigg|_{r=R_C} \end{cases}, \quad (8)$$

where $R_C$ is the boundary between regions $i$ and $j$ ($i, j$ =1-5). The effective parameters are obtained by solving Eq. (7) and Eq. (8) simultaneously under simplified assumptions (the detailed derivation is shown in Supplementary Material Section One 1.1):

$$\frac{u_e}{u_0} \approx \frac{1-SD_1}{1+SD_1}, \frac{k_e}{k_0} \approx \sqrt{1+SD_0}\sqrt{\frac{1+SD_1}{1-SD_1}}, \frac{g_e}{g_0} \approx \frac{k_0^2 u_0}{k_e^2 u_e}, \quad (9)$$

where $S = 4f_s/(i\pi k_0^2 R_4^2) = 4/(i\pi R^2 k_0^2) = 4/(ia^2 k_0^2)$. By combining the calculated scattering coefficient $D_m$ with Eq. (9), the effective parameters of the DNM can be obtained. $D_m$ is calculated in the Supplementary Materials (Section One 1.2) with detailed derivation.

As the $D_m$ is complex, solving Eq. (9) becomes nontrivial. A proper simplification that not only avoids multivalued issues but also ensures the accuracy of the results is needed. Due to the complexity of this structure, excessive simplification would lead to significant errors. Therefore, unlike previous works that simplify Bessel functions, our approximation focuses solely on the Hankel function. For long water waves ($kR \ll 1$), a simplified form of $D_m$ is derived in the Supplementary Materials (Section One 1.3) to facilitate the approximation of the effective parameters:

$$D_m \approx -\frac{\frac{u_3 k_3}{u_0 k_0} J_m(k_0 R_4)\frac{[1+\Omega O_2 k_3(R'-R_3)]}{k_3(R'-R_3)+O_2} - J_m'(k_0 R_4)}{i\left[Y_m^1(k_0 R_4) - Y_m^{1\prime}(k_0 R_4)\right]}, \quad (10)$$

where $Y_m$ is the Neumann function, $m \in \{0,1\}$, and

$$\begin{cases} O_2 \approx \frac{u_3 k_3}{u_0 k_0} \frac{J_m(k_0 R_3)[Y_m(k_0 R_2)+Y_m'(k_0 R_2)TO_1] - Y_m(k_0 R_3)[J_m(k_0 R_2)+J_m'(k_0 R_2)TO_1]}{J_m'(k_0 R_3)[Y_m(k_0 R_2)+Y_m'(k_0 R_2)TO_1] - Y_m'(k_0 R_3)[J_m(k_0 R_2)+J_m'(k_0 R_2)TO_1]} \\ O_1 = -\frac{u_0 k_0}{u_1 k_1} \frac{J_0(k_1 R_2)Y_0'(k_1 R_1) - Y_0(k_1 R_2)J_0'(k_1 R_1)}{J_0'(k_1 R_2)Y_0'(k_1 R_1) - Y_0'(k_1 R_2)J_0'(k_1 R_1)} \\ \Omega = \frac{J_m''(k_3 R_3)Y_m'(k_3 R_3) - J_m'(k_3 R_3)Y_m''(k_3 R_3)}{J_m(k_3 R_3)Y_m'(k_3 R_3) - J_m'(k_3 R_3)Y_m(k_3 R_3)} \\ T = \frac{d\,\mathrm{sinc}\left(\frac{dm}{2R_2}\right)}{b\,\mathrm{sinc}\left(\frac{bm}{2R_2}\right)} \end{cases} \quad (11)$$

To maintain fundamental prediction accuracy and omnidirectional negative refraction, the splits should be sufficiently small ($k\Delta/4 \ll 1$), while both the number of splits $N_o$ and the number of gear teeth $N_i$ should exceed 3. As $N_o$ and $N_i$ increase, the predicted DNM effective parameters become more accurate. Since the CPA calculation is based on the ratio of $a/\lambda$, both the analytical and simulation sections of this Paper focus on $a/\lambda$. The analysis is applicable to different experimental and simulation scales.

DNM parameters are set as $h_1 = 0.03 * h_0$, $R_1 = 0.05 * a$, $R_2 = 0.37 * a$, $R_3 = 0.39 * a$, $R_4 = 0.45 * a$, $N_i = N_o = 8$, $\Delta = 0.1 * a$, $\gamma = 0.8$, by solving Eq. (9) and Eq. (10) simultaneously, the effective reduced water depth $u_e$, the effective gravity $g_e$, and the effective wave number $k_e$ can be predicted as in Fig. 2(A-B). When only $u_e$ is negative (at $a/\lambda \in (0.174, 0.182)$) or when only $g_e$ is negative (at $a/\lambda \in (0.191, 0.229)$), the real part of the wavenumber vanishes ($Re(k_e) = 0$), demonstrating that the DNM prohibits wave propagation in these respective frequency ranges. At the range $a/\lambda \in (0.182, 0.191)$ where both $u_e$ and $g_e$ are simultaneously negative, $k_e$ becomes purely real and negative, enabling negative refraction of propagating waves.

Although the effective parameters predicted by CPA can provide a relatively accurate estimate of the negative refraction range, the specific frequency band still needs to be combined with the energy band structure. The energy band curves and dispersion surface of DNM can be obtained through numerical simulations by COMSOL Multiphysics, shown in Fig. 2(C-D). From the band structure, the bandgap is observed in the range $a/\lambda \in (0.188, 0.2174)$ (gray regions in Figs. 2(A-C)), while the all-angle negative refraction region occurs at $a/\lambda \in (0.1831, 0.188)$ (yellow regions in Figs. 2(A-C)). The outward expansion of isofrequency contours with decreasing frequency (Fig. 2(E)) confirms the persistence of negative refraction. The isofrequency contours show that the DNM deviates from ideal isotropy, likely due to the insufficient numbers of inner ($N_i$) and outer ($N_o$) splits. Without losses of generality, present study focuses primarily on $N_i, N_o = 8$. In the green region of Figs. 2(A-C) ($a/\lambda \in (0.174, 0.1831)$), the dispersion surfaces and equifrequency contours in Figs. 2(D-E) are observed to exhibit discontinuities, indicating that the DNM cannot support propagating waves for all directions. The propagation properties of DNM predicted via $k_e$ exhibit precise alignment with the simulated band structure, evidenced by the consistent color regions across Figs. 2(A-C).

Notably, the operating frequency band is closely linked to variations in structural parameters. When the split size $\Delta$ is increased, both the negative effective water depth and negative effective gravity bands shift to higher frequencies, whereas they shift to lower frequencies when $\Delta$ is decreased. Similarly, moving the split tubes away from center (increasing $R_3$ and $R_4$) shifts the negative water depth band to higher frequencies, while moving tubes toward center shifts it to lower frequencies. By integrating split tubes with a mechanical device to control their outward (increasing $\Delta, R_3, R_4$) or inward (decreasing $\Delta, R_3, R_4$) motion, DNM can enable water wave isolation, negative refraction, and normal transmission at target frequencies.

To evaluate the predicted negative refraction in double-negative frequency ranges and wave isolation in single-negative frequency ranges, specific frequencies are selected from Fig. 2(E) for simulation. The propagation of a plane wave passing through a DNM array composed of 17x17 units in an isosceles triangle region is simulated in Figure 3(A-D), corresponding to $a/\lambda = 0.1847, 0.18563, 0.1875, 0.193$, respectively. In the double-negative frequency range, negative refraction is observed in Fig. 3(A-C), where the wave is deflected to the right by the DNM. As the frequency increases, the negative refraction angle generally decreases. The precise refraction angles at $a/\lambda = 0.18470, 0.18563, 0.18750$ can be calculated by the isofrequency dispersion contours (Fig. 2(E)), with the corresponding refraction angles of $\theta_t = -59.47, -45, -12.7$, respectively. The detailed calculation process can be found in the Supplementary Materials (Section 1.4). As the frequency approaches $a/\lambda = 0.193$ (within the single-negative), the DNM completely blocks transmission, isolating the incident water waves, shown in Fig. 3(D). The simulations reveal a continuous decrease in the negative refraction angle as frequency increases, transitioning smoothly to complete wave blocking, which clearly visualizes the propagation properties of DNM shifting from double-negative effective parameters ($u_e$ and $g_e$ both negative, enabling negative refraction) to single-negative parameters ($g_e$ negative, inducing isolation).

To validate all-angle negative refraction and imaging capabilities, we simulate a point source ($a/\lambda = 0.186$) placed at a distance of $0.75a$ from a slab of the DNM array, shown in Figs. 4(A-B). Without angular losses, Figs. 4(A) demonstrates all-angle imaging, confirming that DNM supports all-angle negative refraction. The energy flow distribution $|\eta|^2$ in Fig. 4(B) clearly shows the surface waves excited at the interfaces, indicating evanescent wave enhanced by DNM. This ability to support surface waves serves as another evidence of negative propagation parameters $u_e$ and $g_e$. These amplified surface waves can achieve subwavelength resolution by enhancing evanescent components beyond the diffraction limit. The magnitude of normalized displacement $|\eta_n|$ after imaging is shown in Fig. 4(C), and the full width at half maximum in the image is $\delta = 2.1a$, with wavelength $\lambda = 5.38a$, satisfying the requirement of subwavelength imaging $\delta < \lambda/2$. The cases for two-point sources separated by $\lambda/2$ and $\lambda/3$ are presented in Supplementary Material (Section Two). These phenomena demonstrate that the double-negative properties of DNM offer more than just wave-bending effect, they also provide capabilities such as supporting surface waves and facilitating subwavelength imaging.

The simulation results validate the negative propagation properties of water waves in DNM, including the double-negative property (wave bending and all-angle imaging) and the single-negative property (wave isolation), at the predicted frequencies. The distributions of the CPA-predicted effective parameters $u_e$ and $g_e$ directly correspond to precise frequency ranges of these phenomena, providing a quantitative mapping from structural parameters to propagation properties. Based on targeted propagation purposes (e.g., specific operating bands and behaviors), the clear mapping enables inverse design of DNM structures, substantially enhancing adaptability to diverse real-world environments and offering high design flexibility.

**Experimental verification**

Although analytical predictions based on CPA require more outer splits ($N_o$) and gear teeth ($N_i$) for accuracy, experimental structures were simplified to reduce viscous losses and fabrication challenges while preserving key wave manipulation effects. To simplify the structure, three key steps are adopted: (1) reducing the number of outer splits while with a larger split size; (2) applying edge rounding to mitigate viscous losses at sharp edges during fluid flow (with rounding radius kept in an appropriate range to preserve negative refraction effects); and (3) widening the gap between tube and gear ($R_3 - R_2$) to reduce internal viscosity-induced dissipation. Without altering the array unit arrangement, $N_o = 3$ represent the approximate limit for this model. Nevertheless, arranging the dual-split gear structures in an alternating checkerboard pattern (e.g., ABBA-type configuration) can achieve nearly equivalent effects (shown in Fig. 5(A)). The DNM parameters are selected as: $h_1 = 0.02 * h_0$, $R_1 = 0.05 * a$, $R_2 = 0.3 * a$, $R_3 = 0.42 * a$, $R_4 = 0.45 * a$, $N_i = N_o = 2$, $\Delta = 0.1 * a$, $\gamma = 0.8$, edge rounding radius: $0.06 * a$ for gears, $0.0136 * a$ for split tubes. The checkerboard DNM array, as analyzed through its band structure and CPA-predicted effective wave number $k_e$ in Fig. 5(B), stays within the anticipated frequency range and exhibits all-angle negative refraction. The actual negative-refraction frequency range is still consistent with the predicted double-negative frequency range, which conforms the feasibility of the checkerboard dual-split DNM.

However, in actual experiments, we quickly found that limited platform size restricts structural unit length $a$, resulting in a very small split $\Delta$ and shallow $h_1$ atop the inner cylindrical structures. This enhances viscous fluid-solid interactions, leading to severe attenuation of propagating water waves. Additionally, the finite experimental space restricts the number of the setting units of checkerboard DNM, as each periodic unit comprises four structures. Without sufficient number of units, checkerboard DNM would hinder effective wave control. To address this issue, the experimental structure is simplified by enlarging the split size $\Delta$ and $h_1$, and aligning the dual-split DNM units without rotation, forming an oriented dual-split DNM, as shown in Fig. 5(C) (unit cell in Fig. 5(D)). The connecting plates are positioned atop the structures, preventing bottom plate from affecting the experimental background water depth $h_0$. Individual units are immersed in a fluorosilicone-modified acrylate solution, which deposits a hydrophobic layer on the structure surfaces to reduce water-structure surface tension. To facilitate immersion, the units are designed with grooves and protrusions for puzzle-like assembly [Fig. 5(D)], enabling them to be spliced into the desired array afterward. Selected parameters are modified as $h_1 = 6.2[\text{mm}]$, $h_0 = 60[\text{mm}]$, $\Delta = 0.6 * a$, with others consistent with those above (unit cell period $a = 60[\text{mm}]$).

The dispersion surface and isofrequency contours of the dual-split DNM are presented in Figs. 5(E) and 5(F) to clearly illustrate its propagation properties. Without the checkerboard arrangement, the dual-split DNM exhibits strong anisotropy, as evidenced by the concave dispersion surface in Fig. 5(E), where wave propagation is highly dependent on direction. Although the isofrequency contours in Fig. 5(F) show that propagation is unsupported in certain directions, most directions still exhibit double-negative properties. Therefore, owing to the wider split, uncomplicated structure, and ease of implementation, we experimentally employ dual-split DNM to verify its negative refraction functionality through a 5x5 triangular array (Fig. 5(C)). At

2.8 Hz (experimental setup detailed in Supplementary Section Three), dual-split DNM induces a bending of the wavefront, as shown in the patterns of *x*-directed velocity field $V$ and displacement distribution $\eta$ [Figs. 6(A-B)]. In this setup where the structure split directly faces the incident wave (Fig. 5(D)), it is striking that after passing the triangular array's edge, the transmitted wave bends sharply to the right—not propagating along the split as expected. This anomalous propagation behavior results from negative refraction functionality of DNM, experimentally demonstrating the feasibility of realizing double-negative properties in practical applications. The experimental results exhibit a -45° negative refraction angle, which is in good agreement with the simulations at 3 Hz, as illustrated in Figs. 6(C-D). The working frequency decreases by 6.7% in experiment, likely due to viscous losses in the experimental environment. The successful small-scale experiment suggests that DNMs with stricter parameters can be experimentally verified under large-scale conditions.

**Discussion**

We present a novel water-wave metamaterial comprising periodically arranged gear-in-split-tube units, marking the first realization of a Veselago-Pendry double negative metamaterial (DNM) for water waves. Based on coherent potential approximation (CPA), its effective parameters (reduced water depth $u_e$ and gravity $g_e$) were accurately predicted via a novel refined scattering approximation. Full-wave simulations validate the DNM's negative propagation properties, revealing distinct propagation behaviors across different frequency ranges, *i.e.*, tunable wave bending and all-angle imaging with surface waves excitation in double-negative bands, and complete wave isolation in single-negative bands. Experimentally, a triangular array of simplified DNM units induces -45° wavefront bending at 2.8 Hz, directly validating negative refraction phenonmenon. Despite a frequency shift of 6.7% caused by viscous losses, the experimental results agree well with simulations, demonstrating its applicability in practical scenarios. By tuning the geometrical parameters of the structure mechanically, the propagation of water waves can be switched among total reflection, negative refraction, and normal refraction at target frequencies. DNMs have potential applications in calming harbors, reducing river-band erosion, and concentrating wave energy, which offer a versatile suite of coastal-engineering tools. The isotropic structural composition shows this DNM structure can also be applied to other wave fields, such as electromagnetic waves and acoustics waves.

**MATERIALS AND METHODS**
**Numerical simulation**

The numerical simulations were based on the Partial Differential Equation (PDE) module of finite-element-method simulation software COMSOL Multiphysics. The water waves are manipulated by the nonlinear water wave equation $\nabla \cdot (u\nabla p) + \frac{\omega^2}{g}p = 0$, where $p = \rho g \eta$, and $u = [\tanh(kh)]/k$. The background water depth $h_0$ is $60\ mm$, with the relationship between frequency and wavelength satisfying $\omega = \sqrt{guk}$.

## Sample fabrication
The experimental sample is made of photosensitive resin through 3D printed.

## Experimental measurement
Wave fields were captured with Fast Fourier Demodulation (FCD) (shown in Fig. S4). The wave-induced distortion of a checkerboard background is recorded and demodulated. The surface elevation η and velocity field V across the entire DNM region can be reconstructed through comparison with reference images. Details are provided in Section Three of the Supplementary Materials.

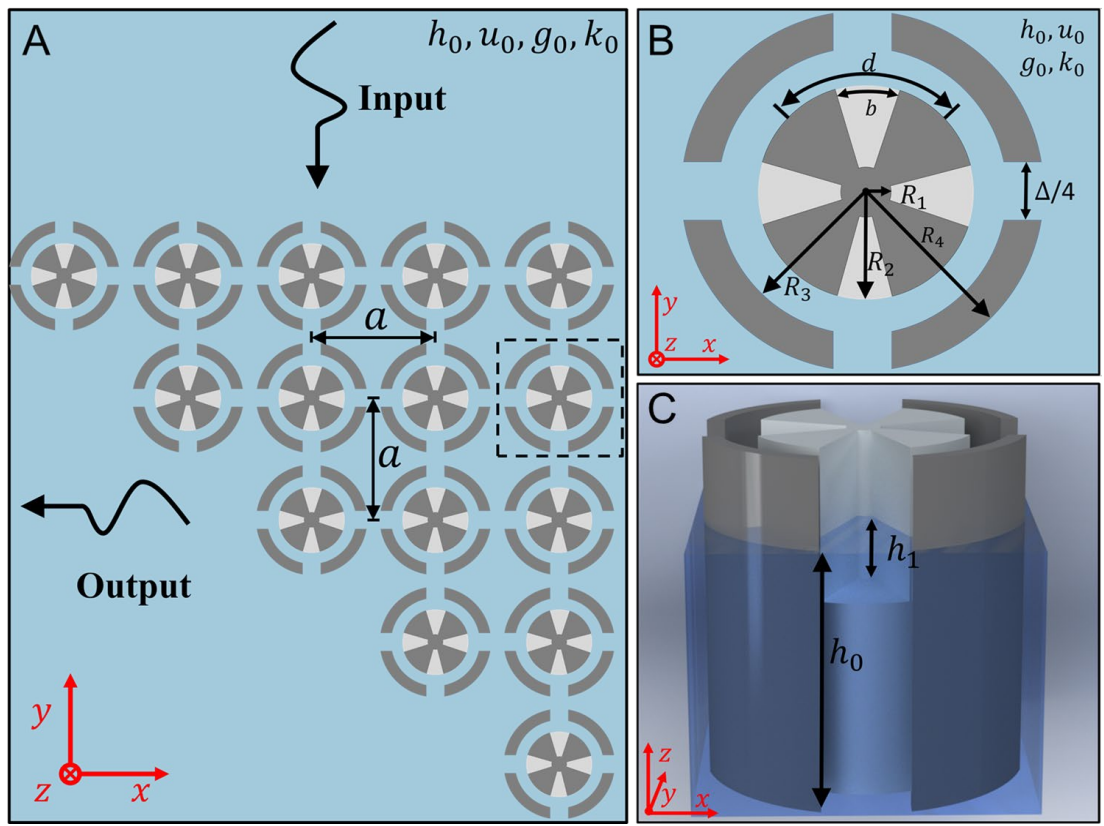

**Fig. 1. Schematic diagram of double-negative metamaterial. (A)** Schematic diagram of DNM 90° twisting water wave direction. Top view **(B)** and side view **(C)** of a single nested ring-and-gear unit.

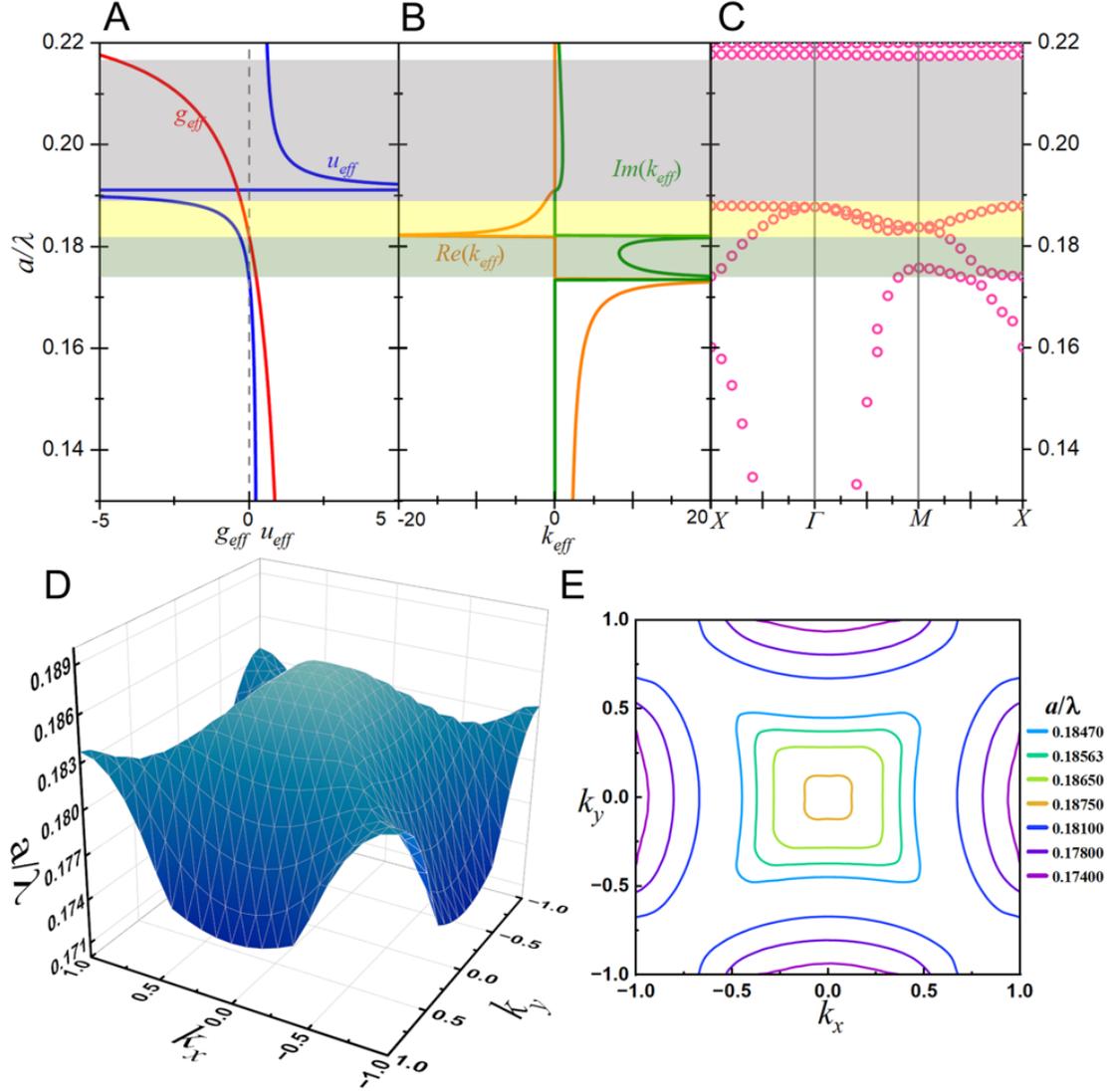

Fig. 2. In the range $0.13 \leq a/\lambda \leq 0.22$, (A) the effective gravity $g_e$ (red line) and the effective reduced water depth $u_e$ (blue line); (B) the real (orange line) and imaginary (green line) parts of the effective wavenumber $k_e$; (C) the band structure of DNM, where the gray shadow region represents the omnidirectional bandgap, the yellow shadow region represents the negative refraction band, and the green shadow region indicates the partial directional bandgap. (D) illustrates the 3D dispersion surfaces of DNM in the negative refraction band, and (E) shows the isofrequency dispersion contours at different frequencies. DNM parameters are $h_1 = 0.03 * h_0$; $R_1 = 0.05 * a$; $R_2 = 0.37 * a$; $R_3 = 0.39 * a$; $R_4 = 0.45 * a$; $N_i = N_o = 8$; $\Delta = 0.1 * a$; $\gamma = 0.8$.

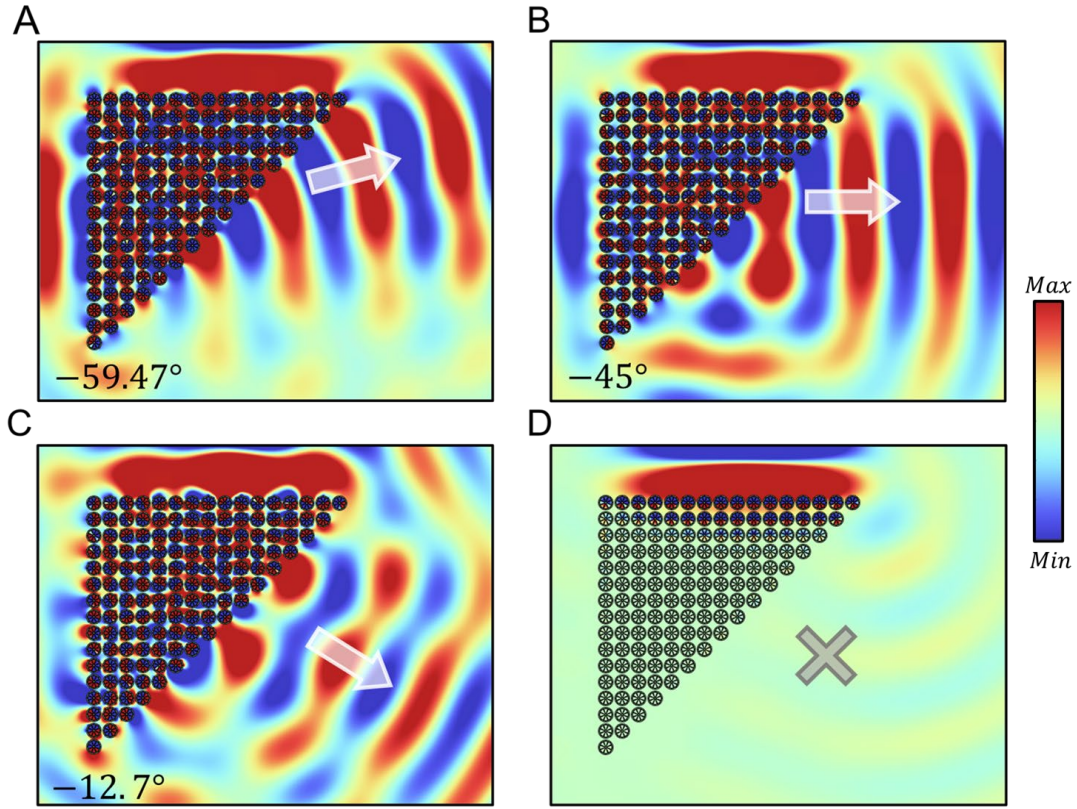

**Fig. 3.** Simulation results of triangle DNM, when the plane wave is incident at $a/\lambda$ is (A) 0.1847, (B) 0.18563, (C) 0.1875, (D) 0.193. DNM parameters are $h_1 = 0.03 * h_0$; $R_1 = 0.05 * a$; $R_2 = 0.37 * a$; $R_3 = 0.39 * a$; $R_4 = 0.45 * a$; $N_i = N_o = 8$; $\Delta = 0.1 * a$; $\gamma = 0.8$.

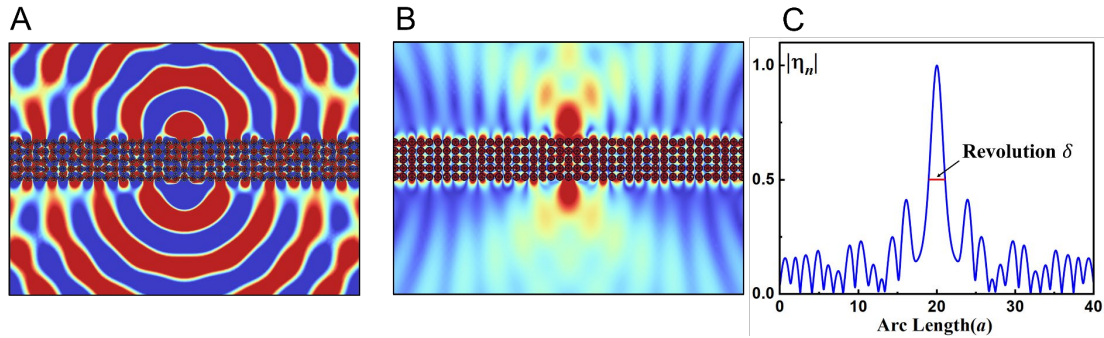

**Fig. 4.** Simulation results of all-angle imaging with a point source positioned at a distance of $0.75a$ from DNM, patterns at $a/\lambda = 0.186$ of (a) vertical displacement $\eta$ and (b) Energy Flow distribution $|\eta|^2$. (c) Magnitude of normalized $|\eta|$ of (a) along the $x$ direction at a distance of $0.75a$ from DNM, which performed over 40 units lengths centered on the source. The red line represents full width at half maximum, which is $1.6a$. DNM parameters are $h_1 = 0.03 * h_0$; $R_1 = 0.05 * a$; $R_2 = 0.37 * a$; $R_3 = 0.39 * a$; $R_4 = 0.45 * a$; $N_i = N_o = 8$; $\Delta = 0.1 * a$; $\gamma = 0.8$.

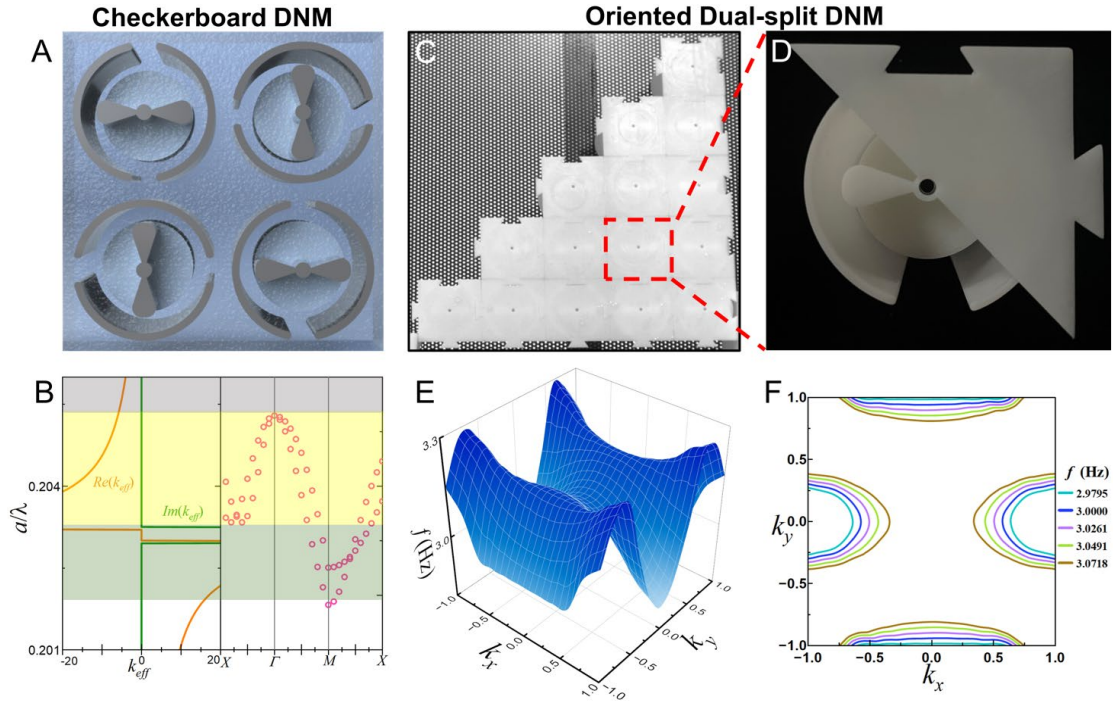

**Fig. 5. (A) Periodic unit cell of the checkerboard DNM and (B) effective parameters and band structure calculated via CPA, with $h_1 = 0.02 * h_0$, and $\Delta = 0.1a$. Experimental sample with $h_1 = 6.2$ [mm], $h_0 = a = 60$ [mm], and $\Delta = 0.6a$ shown in (C) triangular oriented dual-split DNM array and (D) single unit with half the cover hidden. (E) 3D dispersion surfaces of the oriented dual-split DNM in the negative refraction band; (F) isofrequency dispersion contours at various frequencies. Global parameters: $R_1 = 0.05a$, $R_2 = 0.3a$, $R_3 = 0.42a$, $R_4 = 0.45a$, $N_i = N_o = 2$, $\gamma = 0.8$, rounding radii: $0.06a$ for gears and $0.0136a$ for split tubes.**

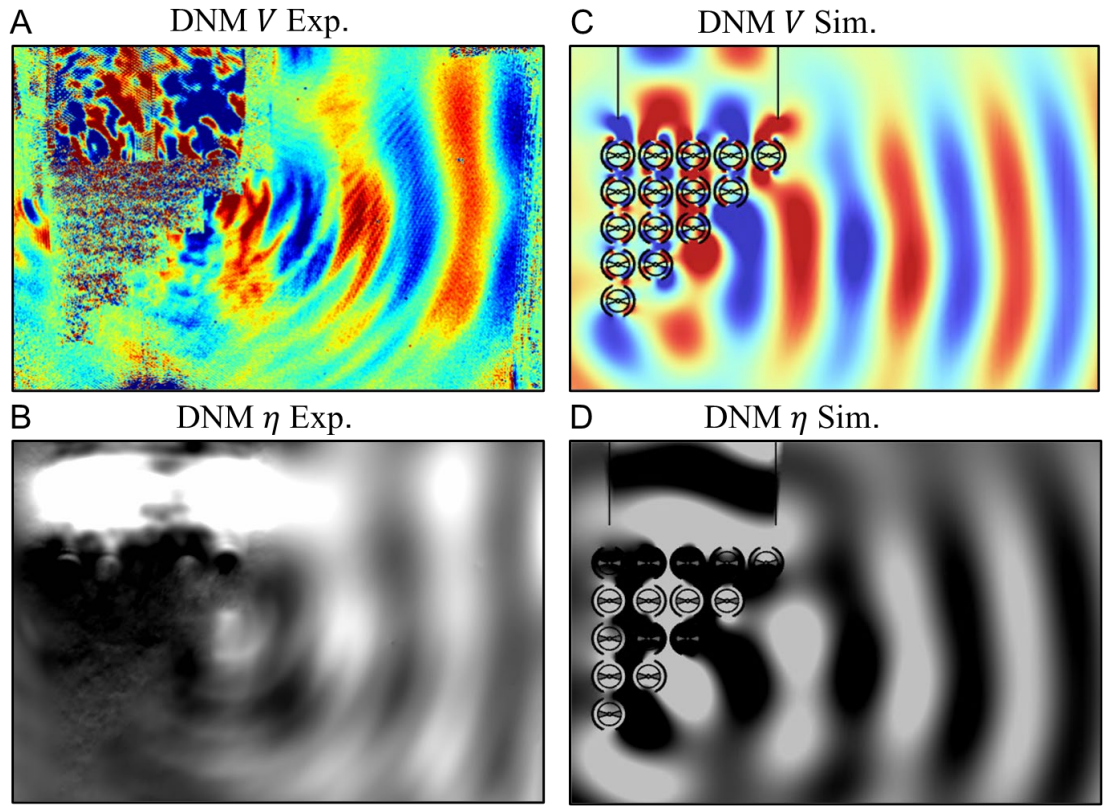

Fig. 6. Experimental results (2.8Hz) of (A) the *x*-directed velocity field $V$ and (B) displacement $\eta$ through the oriented dual-split DNM. Simulation results (3.0Hz) of (C) the *x*-directed velocity field $V$ and (D) displacement $\eta$.

# Supplementary Materials for

# Double Negative Metamaterials in Water Waves

Zixun Ge[1,#], Junke Liao[1,#], Linkang Han[1,#], Qilin Duan[1], Xiaofan Wang[1], Mengwei Dai[2], Shan Zhu[3,*] and Huanyang Chen[1,*]

1. Department of Physics, Xiamen University, Xiamen, 361005, China.

2. Institute of Electromagnetics and Acoustics, School of Electronic Science and Model Microelectronics College, Xiamen University, Xiamen, 361005 China

3. Department of Mechanical Engineering, Guangdong Technion-Israel Institute of Technology, Shantou 515063, China

[#]These authors contributed equally to this work.

Corresponding author. E-mail: *zhushanbay@gmail.com ; kenyon@xmu.edu.cn

**Section One: Theoretical Analysis**

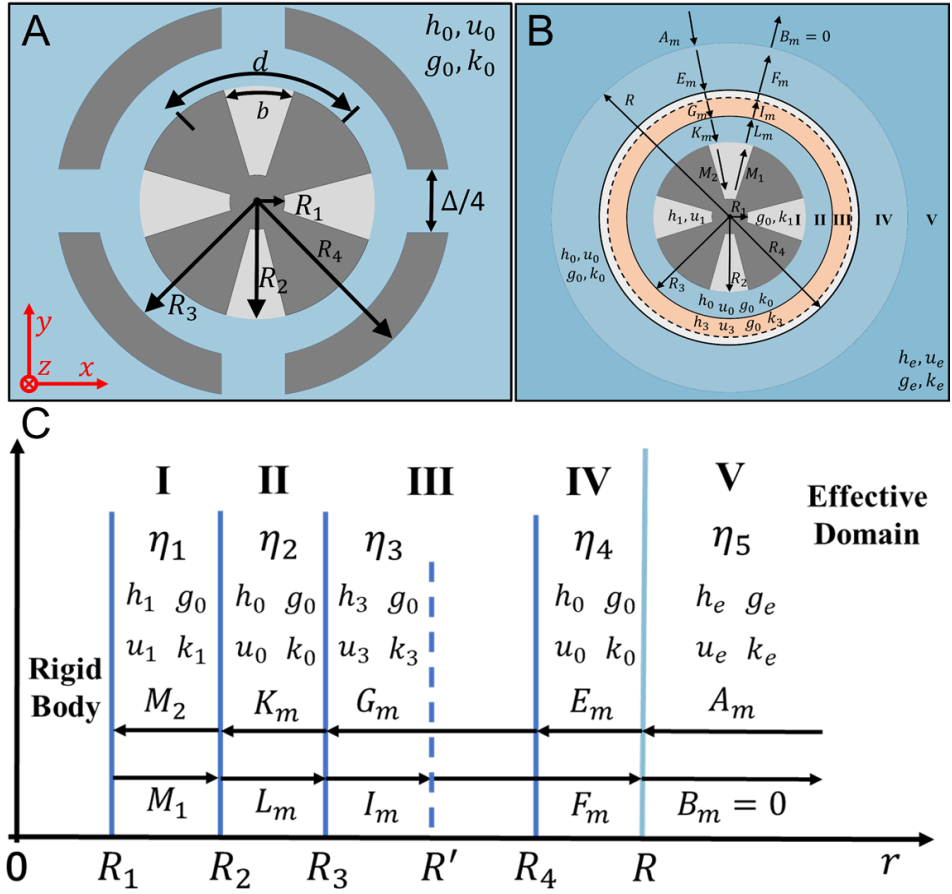

**Fig. S1. Schematic diagram of double-negative metamaterial (DNM)'s unit structure. (A) Actual structural parameter diagram and (B) scattering distribution and propagation parameters distribution of DNM single structure after approximating the split tubes as two effective fluid layers. (C) Distribution diagram of propagation parameters at each region with path length.**

### 1.1 Derivation of Eq. (9) in main text

For periodic arrays with symmetric structure, coherent-potential-approximation (CPA) homogenizes the whole array into a region with effective propagation parameters $(k_e, u_e, g_e, h_e)$. The analysis focuses on the individual unit cell (Fig. S1(A)), where the surrounding DNM beyond radius R is treated as an effective homogeneous fluid, as shown in Fig. S1(B). The filling radio of the periodic arrays can be written as $f_s = \frac{\pi R_4^2}{\pi R^2} = \frac{\pi R_4^2}{a^2}$. In Fig. S1(B), the structure is divided into five regions, i.e., region I, the gap of the central gear with effective parameters $(k_1, u_1, g_0, h_1)$; region II, the interior background between the gear and split tube with effective parameters $(k_0, u_0, g_0, h_0)$; region III, the split tube itself, replaced by an effective-depth shell with effective parameters $(k_3, u_3, g_0, h_3)$; region IV, the exterior background with effective parameters $(k_0, u_0, g_0, h_0)$; and region V is the CPA-homogenized fluid of DNM with effective parameters $(k_e, u_e, g_e, h_e)$. Region V is matched continuously to region IV at the radius $r = R = a/\sqrt{\pi}$, where $a$ is the lattice constant of the periodic arrays. The dispersion of water waves can be written as

$$\omega^2 = gk\tanh(kh), \qquad (S1)$$

where $\omega$ is the angular frequency, $k$ is the wave vector ($k = 2\pi/\lambda$), $\lambda$ is the wavelength, $g$ is the gravity acceleration, and $h$ is the water depth. Define $u = [\tanh(kh)]/k$, and $u$ is the reduced water depth. The vertical displacement $\eta$ of the water surface is obtained by $\eta = \frac{i}{\omega}\frac{\partial}{\partial z}\Phi(x,y,z=0)e^{-i\omega t}$, where $\Phi(x,y,z)e^{-i\omega t}$ is the velocity potential. We analyze the vertical displacement $\eta$ of the water wave in cylindrical coordinates $(r,\phi)$, within regions I to V. Fig.S1 (c) shows that the potential $\eta$ in regions I to V will be

$$\begin{aligned}
\eta_1 &= [M_1 H_0^1(k_1 r) + M_2 H_0^2(k_1 r)] & (R_2 \leq r \leq R_1)\\
\eta_2 &= \sum_m [K_m J_m(k_0 r) + L_m H_m^1(k_0 r)]e^{im\theta} & (R_3 \leq r \leq R_2)\\
\eta_3 &= \sum_m [G_m J_m(k_3 r) + I_m H_m^1(k_3 r)]e^{im\theta} & (R' \leq r \leq R_3), \qquad (S2)\\
\eta_4 &= \sum_m [E_m J_m(k_0 r) + F_m H_m^1(k_0 r)]e^{im\theta} & (R \leq r \leq R_4)\\
\eta_5 &= \sum_m [A_m J_m(k_e r)]e^{im\theta} & (R \leq r)
\end{aligned}$$

where the $m$-order Bessel function $J_m$ and the second kind of 0-order Hankel function $H_0^2$ describe incident waves, and the first kind of $m$-order Hankel function $H_m^1$ represents scattering waves. $k_0$ is the wavenumber of water waves in the background fluid surrounding DNM. There are continuous boundary conditions

$$\begin{cases} p_i(R_C) = p_j(R_C)\\ \frac{iu_i}{\omega\rho}\frac{\partial p_i(r)}{\partial r}\Big|_{r=R_C} = \frac{iu_j}{\omega\rho}\frac{\partial p_j(r)}{\partial r}\Big|_{r=R_C} \end{cases}, \qquad (S3)$$

where $R_C$ is the boundary between region $i$ and $j$ ($i,j$ =I-V), and $p$ is the hydrostatic pressure, defined as $p = \rho g \eta$, with $\rho$ being the fluid density. By matching the wave field at the interface between the CPA-homogenized region V and the outer background region IV, the relationship between the DNM's effective parameters and the structure scattering coefficients can be derived. Combining $\eta_4$ and $\eta_5$ in Eq. (S2) with (S3), the simultaneous equations are

$$\begin{cases} g_e A_m J_m(k_e R) = g_0 [E_m J_m(k_0 R) + F_m H_m^1(k_0 R)]\\ g_e u_e k_e A_m J_m'(k_e R) = g_0 u_0 k_0 [E_m J_m'(k_0 R) + F_m H_m^{1\,\prime}(k_0 R)] \end{cases}. \qquad (S4)$$

We can obtain the relation between effective parameters ($k_e, u_e$) and $D_m$:

$$D_m = -\frac{u_0 k_0 J_m'(k_0 R) - J_m(k_0 R)[u_e k_e J_m'(k_e R)/J_m(k_e R)]}{u_0 k_0 H_m^{1\,\prime}(k_0 R) - H_m^1(k_0 R)[u_e k_e J_m'(k_e R)/J_m(k_e R)]}. \qquad (S5)$$

Since the unknown quantity $k_e$ in Eq. (S5) is the variable of the nonlinear Bessel functions $J_m'(k_e R)$ and $J_m(k_e R)$, the solution will yield multivalued results. As the water wavelength is much larger than the size of structure, the equation can be simplified. Following the simplification methods of Bessel and Hankel functions shown in Ref. (1), Eq. (S5) can be reduced as

$$D_0 \approx -\frac{u_0 k_0\left(-\frac{k_0 R}{2}\right) - 1\left[\frac{u_e k_e\left(-\frac{k_e R}{2}\right)}{1}\right]}{u_0 k_0\left(-\frac{2}{i\pi k_0 R}\right) - 0} = -\frac{1}{S}\left(1 - \frac{u_e k_e^2}{u_0 k_0^2}\right), \qquad (S6)$$

$$D_1 \approx -\frac{\frac{u_0 k_0}{2} - \frac{k_0 R}{2}\left[\frac{u_e}{R}\right]}{u_0 k_0 2i/(\pi(k_0 R)^2) - \frac{2}{i\pi k_0 R}\left[\frac{u_e}{R}\right]} = \frac{1}{S}\frac{u_0 - u_e}{u_0 + u_e}, \tag{S7}$$

where $S = 4f_s/(i\pi k_0^2 R_4^2) = 4/(i\pi R^2 k_0^2) = 4/(ia^2 k_0^2)$. From Eq. (S6) and (S7), we can derive the equation showing in main text:

$$\frac{u_e}{u_0} = \frac{1 - SD_1}{1 + SD_1}, \tag{S8}$$

$$\frac{k_e}{k_0} = \sqrt{1 + SD_0}\sqrt{\frac{1 + SD_1}{1 - SD_1}}. \tag{S9}$$

From (S8) and (S9), the following equations can be:

$$\frac{g_e}{g_0} = \frac{k_0^2 u_0}{k_e^2 u_e}, \quad \frac{h_e}{h_0} = \frac{k_0 \mathrm{atanh}(u_e k_e)}{k_e \mathrm{atanh}(u_0 k_0)}. \tag{S10}$$

Then the effective parameters are obtained.

### 1.2 Derivation of Eq. (10) in main text

After establishing the relation between the scattering coefficients of gear-in-split-tube structure and the effective parameters of DNM, the scattering coefficients $D_m$ can be determined. Starting from the inner region, the flow will not pass through the hard rigid body at $r = R_1$, $\frac{iu_1}{\omega\rho}\frac{\partial p_1(r)}{\partial r}|_{r=R_1} = 0$. The equation can be shown as

$$M_1 k_1 H_0^{1'}(k_1 R_1) + M_2 k_1 H_0^{2'}(k_1 R_1) = 0. \tag{S11}$$

Based on dispersion relationship Eq. (S1), $k_1$ can be calculated by

$$g_0 k_0 \tanh(k_0 h_0) = g_0 k_1 \tanh(k_1 h_1). \tag{S12}$$

with unchanged gravity $g_0$. When $r = R_2$, the boundary conditions can be divided into two parts. One is the continuous condition for $0 < |\theta| < b/2R_2$, satisfying Eq.(S3),

$$\begin{cases} g_0 \sum_m [K_m J_m(k_0 R_2) + L_m H_m^1(k_0 R_2)]e^{im\theta} = g_0[M_1 H_0^1(k_1 R_2) + M_2 H_0^2(k_1 R_2)] \\ u_0 g_0 k_0 \sum_m [K_m J_m'(k_0 R_2) + L_m H_m^{1'}(k_0 R_2)]e^{im\theta} = u_1 g_0 k_1 [M_1 H_0^{1'}(k_1 R_2) + M_2 H_0^{2'}(k_1 R_2)]' \end{cases}$$
$$\tag{S13}$$

where $g_0$ keeps the same and the prime symbol (') denotes the derivative of the function. The other one is hard boundary condition for $b/2R_2 < |\theta| < d/2R_2$, which can be derived from $\frac{iu_0}{\omega\rho}\frac{\partial p_2(r)}{\partial r}|_{r=R_2} = 0$ as

$$\sum_m k_0 [K_m J_m'(k_0 R_2) + L_m H_m^{1'}(k_0 R_2)]e^{im\theta} = 0. \tag{S14}$$

Eq. (S13) represents the boundary continuity condition between the outer gear region (supporting multiple modes) and the inner gear region (supporting only the fundamental mode). This equation is challenging to be solved, as it involves matching a modal series expansion with a single-mode expression. According to ref. (2), we can assume that the response of the system is dominated by a single $m$-mode excited in Region II. Thus, the mode superposition on the left-hand side of Eq. (S13) can be approximated as including only one term corresponding to the excited $m$-mode. Since the right-hand side contains only the fundamental mode, each *m*-mode on the left-hand side can be represented by the fundamental mode on the right-hand side multiplied by a scaling factor $\alpha$. $r = R_2$, for $0 < |\theta| < b/2R_2$, Eq. (S9) can be written as:

$$\begin{cases} \sum_m [K_m J_m(k_0 R_2) + L_m H_m^1(k_0 R_2)] e^{im\theta} = \sum_m \alpha_m [M_1 H_0^1(k_1 R_2) + M_2 H_0^2(k_1 R_2)] \\ \frac{u_0 k_0}{u_1 k_1} \sum_m [K_m J_m'(k_0 R_2) + L_m H_m^{1\prime}(k_0 R_2)] e^{im\theta} = \sum_m \alpha_m [M_1 H_0^{1\prime}(k_1 R_2) + M_2 H_0^{2\prime}(k_1 R_2)]' \end{cases}$$

(S15)

where the accumulation sign can be removed:

$$\begin{cases} [K_m J_m(k_0 R_2) + L_m H_m^1(k_0 R_2)] e^{im\theta} = \alpha_m [M_1 H_0^1(k_1 R_2) + M_2 H_0^2(k_1 R_2)] \\ \frac{u_0 k_0}{u_1 k_1} [K_m J_m'(k_0 R_2) + L_m H_m^{1\prime}(k_0 R_2)] e^{im\theta} = \alpha_m [M_1 H_0^{1\prime}(k_1 R_2) + M_2 H_0^{2\prime}(k_1 R_2)]' \end{cases}$$

(S16)

where $\sum_m \alpha_m = 1$. For $b/2R_2 < |\theta| < d/2R_2$, Eq. (S14) can be simplified to:

$$[K_m J_m'(k_0 R_2) + L_m H_m^{1\prime}(k_0 R_2)] e^{im\theta} = 0 \qquad (S17)$$

Due to the symmetry of the structure, only a single period of the gear need to be analyzed. Integrating Eq. (S16) over the range $0 < |\theta| < b/2R_2$ leads to the relation that

$$[K_m J_m(k_0 R_2) + L_m H_m^1(k_0 R_2)] \operatorname{sinc}\left(\frac{bm}{2R_2}\right) = \alpha_m [M_1 H_0^1(k_1 R_2) + M_2 H_0^2(k_1 R_2)].$$

(S18)

Integrating Eq. (S16) and (S17) over the range $0 < |\theta| < d/2R_2$, here is:

$$[K_m J_m'(k_0 R_2) + L_m H_m^{1\prime}(k_0 R_2)] \operatorname{sinc}\left(\frac{dm}{2R_2}\right) \frac{d}{b} = \frac{u_1 k_1}{u_0 k_0} \alpha_m [M_1 H_0^{1\prime}(k_1 R_2) + M_2 H_0^{2\prime}(k_1 R_2)].$$

(S19)

Combining the Eq.(S11), (S18) and (S19), the ratio of coefficients can be obtained by

$$\frac{L_m}{K_m} = -\frac{J_m(k_0 R_2) + J_m'(k_0 R_2) T(b,d,R_2,m) O_1}{H_m^1(k_0 R_2) + H_m^{1\prime}(k_0 R_2) T(b,d,R_2,m) O_1}, \qquad (S20)$$

where $T(b,d,R_2,m) = d\operatorname{sinc}\left(\frac{dm}{2R_2}\right) / \left[b\operatorname{sinc}\left(\frac{bm}{2R_2}\right)\right]$, $O_1 = \frac{u_0 k_0}{u_1 k_1} \frac{H_0^1(k_1 R_2) H_0^{2\prime}(k_1 R_1) - H_0^2(k_1 R_2) H_0^{1\prime}(k_1 R_1)}{H_0^{2\prime}(k_1 R_2) H_0^{1\prime}(k_1 R_1) - H_0^{1\prime}(k_1 R_2) H_0^{2\prime}(k_1 R_1)}$. Since $\alpha_m$ cancels out in the derivation, it plays no role in determining the result.

    For the studied region II, III and IV, it becomes necessary to consider how water waves passing through the rigid split tube. When the wavelength of water waves is much longer than the slit width ($k\Delta/4 \ll 1$), the rigid split tube can be replaced by an effective two-layer shell (see the orange and gray regions in Fig. S1(B)) (*3*). The inner layer ($R_3 < r < R'$) is an effective fluid of uniform water depth and is continuous with region II at $r = R_3$. The reduced water depth of effective fluid is $u_3 = u_0/n_{st}^2$ with a thickness of $(R_4 - R_3)/n_{st}$, where $n_{st} = 2\pi R_4/\Delta$ (*3*). The boundary is at $R' = R_3 + (R_4 - R_3)/n_{st}$. The wave number is $k_3 = n_{st} k_0$ and the gravity acceleration is unchanged. The outer layer satisfies field continuity at both the inner

interface ($r = R'$) and the outer interface ($r = R_4$), maintaining the consistency of the total wall thickness (shown in Fig. S1(B)). Considering that the fields at two surfaces ($r = R_4, R'$) of the outer layer matches, the continuous condition between regions III and IV are $p_3(R') = p_4(R_4)$ and $\frac{iu_3}{\omega\rho}\frac{\partial p_3(r)}{\partial r}|_{r=R'} = \frac{iu_0}{\omega\rho}\frac{\partial p_4(r)}{\partial r}|_{r=R_4}$, while the continuity between Regions II and III can be directly derived from Eq. (S3). Then the equations can be derived as

$$\begin{cases} G_m J_m(k_3 R_3) + I_m H_m^1(k_3 R_3) = K_m J_m(k_0 R_3) + L_m H_m^1(k_0 R_3) \\ u_3 k_3 [G_m J_m'(k_3 R_3) + I_m H_m^{1'}(k_3 R_3)] = u_0 k_0 [K_m J_m'(k_0 R_3) + L_m H_m^{1'}(k_0 R_3)] \\ E_m J_m(k_0 R_4) + F_m H_m^1(k_0 R_4) = G_m J_m(k_3 R') + I_m H_m^1(k_3 R') \\ u_0 k_0 [E_m J_m'(k_0 R_4) + F_m H_m^{1'}(k_0 R_4)] = u_3 k_3 [G_m J_m'(k_3 R') + I_m H_m^{1'}(k_3 R')] \end{cases}$$

(S21)

The ratios of coefficients are calculated as,

$$\frac{I_m}{G_m} = -\frac{J_m(k_3 R_3) - J_m'(k_3 R_3) O_2}{H_m^1(k_3 R_3) - H_m^{1'}(k_3 R_3) O_2},$$ (S22)

$$O_2 = \frac{u_3 k_3}{u_0 k_0}\frac{J_m(k_0 R_3) + \frac{L_m}{K_m} H_m^1(k_0 R_3)}{J_m'(k_0 R_3) + \frac{L_m}{K_m} H_m^{1'}(k_0 R_3)},$$ (S23)

$$D_m = \frac{F_m}{E_m} = -\frac{J_m(k_0 R_4) - J_m'(k_0 R_4) O_3}{H_m^1(k_0 R_4) - H_m^{1'}(k_0 R_4) O_3},$$ (S24)

$$O_3 = \frac{u_0 k_0}{u_3 k_3}\frac{J_m(k_3 R') + \frac{I_m}{G_m} H_m^1(k_3 R')}{J_m'(k_3 R') + \frac{I_m}{G_m} H_m^{1'}(k_3 R')}.$$ (S25)

Based on Eq. (S20), (S22) and (S24), the scattering coefficient $D_m$ is formulated:

$$D_m = \frac{F_m}{E_m} = -\frac{J_m(k_0 R_4) - J_m'(k_0 R_4) O_3}{H_m^1(k_0 R_4) - H_m^{1'}(k_0 R_4) O_3},$$ (S26)

$$\begin{cases} O_3 = \frac{u_0 k_0}{u_3 k_3}\frac{J_m(k_3 R') H_m^1(k_3 R_3) - J_m(k_3 R') H_m^{1'}(k_3 R_3) O_2 - [J_m(k_3 R_3) H_m^1(k_3 R') - J_m'(k_3 R_3) H_m^1(k_3 R') O_2]}{J_m'(k_3 R') H_m^1(k_3 R_3) - J_m'(k_3 R') H_m^{1'}(k_3 R_3) O_2 - [J_m(k_3 R_3) H_m^{1'}(k_3 R') - J_m'(k_3 R_3) H_m^{1'}(k_3 R') O_2]} \\ O_2 = \frac{u_3 k_3}{u_0 k_0}\frac{J_m(k_0 R_3) H_m^1(k_0 R_2) + J_m(k_0 R_3) H_m^{1'}(k_0 R_2) T(b,d,R_2,m) O_1 - [J_m(k_0 R_2) H_m^1(k_0 R_3) + J_m'(k_0 R_2) H_m^1(k_0 R_3) T(b,d,R_2,m) O_1]}{J_m'(k_0 R_3) H_m^1(k_0 R_2) + J_m'(k_0 R_3) H_m^{1'}(k_0 R_2) T(b,d,R_2,m) O_1 - [J_m(k_0 R_2) H_m^{1'}(k_0 R_3) + J_m'(k_0 R_2) H_m^{1'}(k_0 R_3) T(b,d,R_2,m) O_1]} \\ O_1 = \frac{u_0 k_0}{u_1 k_1}\frac{H_0^1(k_1 R_2) H_0^{2'}(k_1 R_1) - H_0^2(k_1 R_2) H_0^{1'}(k_1 R_1)}{H_0^{2'}(k_1 R_2) H_0^{1'}(k_1 R_1) - H_0^{1'}(k_1 R_2) H_0^{2'}(k_1 R_1)} \\ T(b,d,R_2,m) = \frac{d\,\text{sinc}\left(\frac{dm}{2R_2}\right)}{b\,\text{sinc}\left(\frac{bm}{2R_2}\right)} \end{cases}$$

(S27)

### 1.3. Simplified equations

In Section One 1.1, to derive the relationship between $D_m$ and the effective parameters ($u_e, g_e$), we have adopted a commonly approximate treatment of the Bessel functions $J_m(x)$ and Hankel funtcions $H_m(x)$ (*1*). When all Bessel functions are approximated as simplified linear functions, the nonlinear multi-valued problems of the scattering functions can be neglected. This simplification indeed avoids the complexity of simultaneously dealing with both real and imaginary parts of Hankel function, significantly accelerating the computation process. However, the approximation method adopted in ref. (*2*) should not be employed to simplify the analytical expression of $D_m$ (Eq.(S26)), since it leads to unreliable results when applying the reduced $D_m$ to solve Eq. (S8) and (S10) for effective parameters. Unlike Eq. (S5) used for simplification in Section One 1.1, $D_m$ (Eq.(S26)) involves numerous Bessel functions. In such complex computations with Bessel functions, the simplification method from Ref. (*2*) would give rise to severe oversimplification, resulting in substantial deviation from the exact solution. Therefore, a novel approach which can avoid the multivalued issues while preserving the accuracy is required.

As $D_m$ serves as a constant without unknown variables, we retain the full forms of the Bessel functions $J_m(x)$ and $Y_m(x)$ to ensure analytical accuracy. The Hankel function $H_m(x)$ can be simplified to $H_m(x) \approx i * Y_m(x)$ ($m \in \{0,1\}$) to avoid the multivalued issues associated with complex analysis. This simplification is valid when the argument of the Hankel function $H_m(x)$ satisfies $x \ll 1$. It is worth noting that not all Hankel functions require approximation. The Hankel functions in $O_1$ can be directly simplified after expansion ($H_m^n(kR) = J_m^n(kR) + iY_m^n(kR)$):

$$O_1 = -\frac{u_0 k_0}{u_1 k_1} \frac{J_0(k_1 R_2)Y_0'(k_1 R_1) - Y_0(k_1 R_2)J_0'(k_1 R_1)}{J_0'(k_1 R_2)Y_0'(k_1 R_1) - Y_0'(k_1 R_2)J_0'(k_1 R_1)}. \quad \text{(S28)}$$

Furthermore, to enhance solution accuracy and simplify computations, the analytical expression $O_3$ should be simplified, which is:

$$O_3 = \frac{u_0 k_0}{u_3 k_3} \frac{J_m(k_3 R')H_m^1(k_3 R_3) - H_m^1(k_3 R')J_m(k_3 R_3) - [J_m(k_3 R')H_m^{1\,\prime}(k_3 R_3) - H_m^1(k_3 R')J_m'(k_3 R_3)]O_2}{J_m'(k_3 R')H_m^1(k_3 R_3) - H_m^{1\,\prime}(k_3 R')J_m(k_3 R_3) - [J_m'(k_3 R')H_m^{1\,\prime}(k_3 R_3) - H_m^{1\,\prime}(k_3 R')J_m'(k_3 R_3)]O_2}.$$

(S29)

Based on (*4*) and (*5*), with $d = k_3(R' - R_3); x = k_3 R_3$, the $O_3$ would be transformed to

$$\begin{cases} O_3^{(0)} \approx \frac{u_0 k_0}{u_3 k_3} \frac{d[J_0(x)Y_0'(x) - J_0'(x)Y_0(x)] + [(J_0(x)Y_0'(x) - J_0'(x)Y_0(x))]O_2^{(0)}}{J_0(x)Y_0'(x) - J_0'(x)Y_0(x) + [d[J_0''(x)Y_0'(x) - J_0'(x)Y_0''(x)]]O_2^{(0)}} \\ O_3^{(1)} \approx \frac{u_0 k_0}{u_3 k_3} \frac{d[J_1(x)Y_1'(x) - J_1'(x)Y_1(x)] + [(J_1(x)Y_1'(x) - J_1'(x)Y_1(x))]O_2^{(1)\prime}}{J_1(x)Y_1'(x) - J_1'(x)Y_1(x) + [d[J_1''(x)Y_1'(x) - J_1'(x)Y_1''(x)]]O_2^{(1)}} \end{cases} \quad \text{(S30)}$$

As a result, we arrive at the approximated expression for $D_m$

$$D_m \approx -\frac{\frac{u_3 k_3}{u_0 k_0} J_m(k_0 R_4) \frac{[1+\Omega O_2 k_3(R'-R_3)]}{k_3(R'-R_3)+O_2} - J_m'(k_0 R_4)}{i[Y_m^1(k_0 R_4) - Y_m^{1\,\prime}(k_0 R_4)]} \quad \text{(S31)}$$

Where $m \in \{0,1\}$, $\Omega = \frac{J_m''(k_3 R_3)Y_m'(k_3 R_3) - J_m'(k_3 R_3)Y_m''(k_3 R_3)}{J_m(k_3 R_3)Y_m'(k_3 R_3) - J_m'(k_3 R_3)Y_m(k_3 R_3)}$, and

$$\begin{cases} O_2 \approx \frac{u_3 k_3}{u_0 k_0} \frac{J_m(k_0 R_3)[Y_m(k_0 R_2) + Y_m'(k_0 R_2)TO_1] - Y_m(k_0 R_3)[J_m(k_0 R_2) + J_m'(k_0 R_2)TO_1]}{J_m'(k_0 R_3)[Y_m(k_0 R_2) + Y_m'(k_0 R_2)TO_1] - Y_m'(k_0 R_3)[J_m(k_0 R_2) + J_m'(k_0 R_2)TO_1]} \\ O_1 = -\frac{u_0 k_0}{u_1 k_1} \frac{J_0(k_1 R_2)Y_0'(k_1 R_1) - Y_0(k_1 R_2)J_0'(k_1 R_1)}{J_0'(k_1 R_2)Y_0'(k_1 R_1) - Y_0'(k_1 R_2)J_0'(k_1 R_1)} \\ T = \frac{d\,\text{sinc}\left(\frac{dm}{2R_2}\right)}{\left[b\,\text{sinc}\left(\frac{bm}{2R_2}\right)\right]} \end{cases} \quad \text{. (S32)}$$

## 1.4. Negative refraction angle calculation

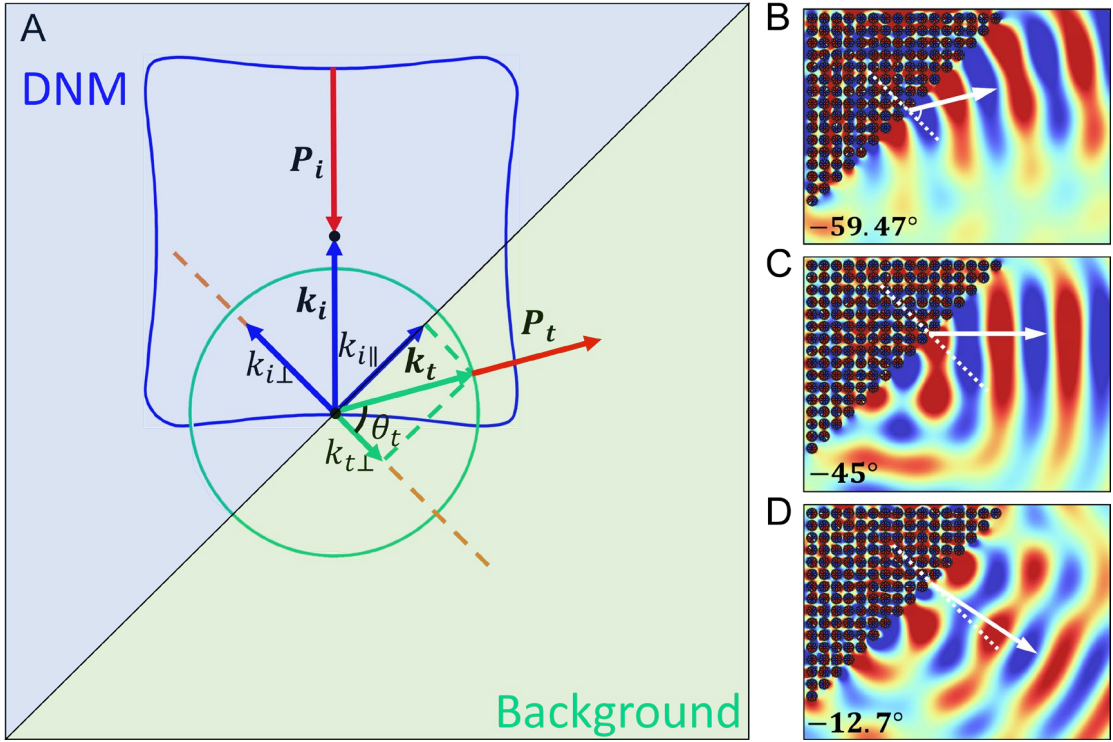

**Fig. S2 (A) Wave vectors and energy flux $P_i$ for a wave incident from the triangular DNM into the background. The blue, nearly rectangular contour is the isofrequency contour of DNM, while the green circle is that of the background. (B–D) Displacement distribution patterns of refraction angles for ratios 0.1847, 0.18563, and 0.1875.**

Figure S2 illustrates the changes of the wave vector and energy flux from incident ($k_i$, $P_i$) to transmitted ($k_t$, $P_t$) for the water wave passing through the DNM into the background fluid. In the double-negative region, the wave vector $k_i$ is antiparallel to $P_i$. When the wave reaches the slanted edge of the isosceles triangular slab array, the wave vector $k_i$ can be decomposed into vertical and tangential components with respect to the interface:

$$k_{i\parallel} = k_{i\perp} = \frac{k_i}{\sqrt{2}} \tag{S33}$$

The interface satisfies the continuity of the tangential wave vector,

$$k_{i\parallel} = k_{t\parallel} = \frac{k_i}{\sqrt{2}}, \tag{S34}$$

where $k_{t\parallel}$ is the tangential component of the wave vector $k_t$ of the wave transmitted into the background. As the wave travels from DNM into the background region, the vertical wave-vector component $k_{t\perp}$ reverses direction, switching from opposing the direction of energy flux to aligning with it. The vertical component $k_{t\perp}$ can be obtained from the background isofrequency contour using the transmitted wave vector $k_t$:

$$k_{t\perp} = \sqrt{k_t^2 - k_{t\|}^2}. \tag{S35}$$

In the isotropic background region, the wave vector $k_t$ is aligned with the direction of the energy flux $P_t$. Once the components of the wave vector of the propagating water waves are known, the corresponding propagation direction can be determined, and the refraction angle $\theta_t$ can be calculated as,

$$\theta_t = -90° + \text{atan}\frac{k_{t\perp}}{k_{t\|}} \tag{S36}$$

Based on this method, the refraction angles for the various propagation patterns in the main text can be derived. As shown in the Figs. S2(B-D), when the ratio is 0.18470, 0.18563, 0.18750, the corresponding refraction angles are $\theta_b = -59.47, -45, -12.7$, respectively. The refraction angles predicted by this method agree well with the simulated refraction angles, enabling precise control over the bending of the water waves.

**Section Two: Simulations**

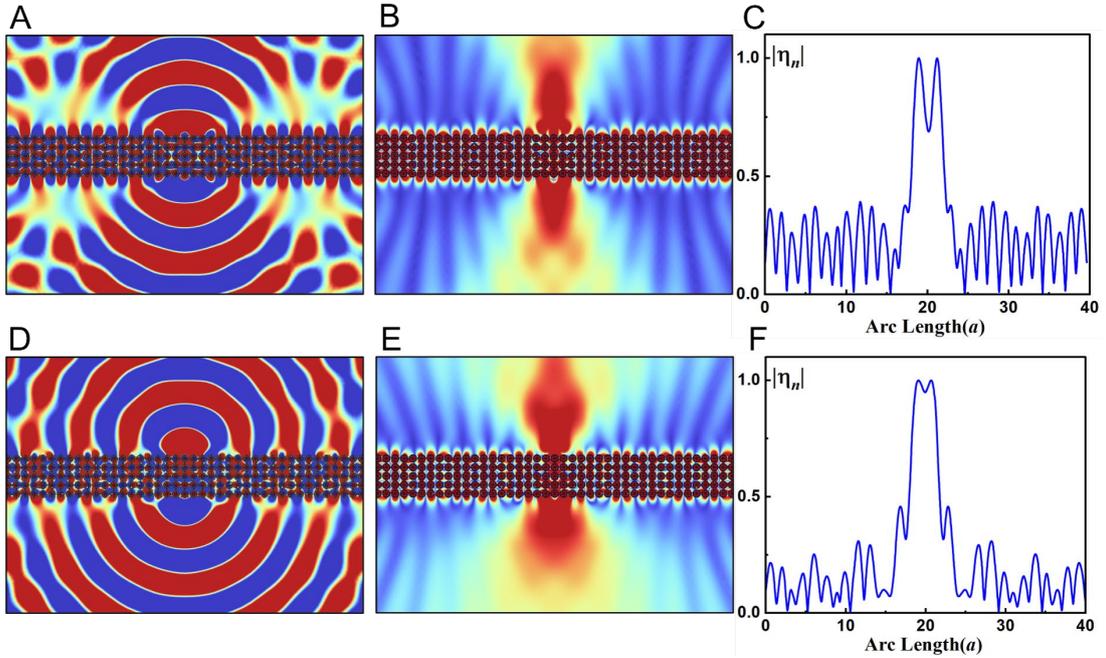

**Fig. S3. Simulation results of all-angle imaging with two-point sources positioned at a distance of $0.75a$ from DNM with a separation of $0.5\lambda$ in (A–C) and $0.33\lambda$ in (D–F). The patterns at $a/\lambda = 0.186$ of (A, D) vertical displacement $\eta$ and (B, E) Energy Flow distribution $|\eta|^2$. (C, F) Normalized $|\eta|$ of (A, D) along the $x$ direction at a distance of $0.75a$ from DNM. DNM parameters are $h_1 = 0.03 * h_0$; $R_1 = 0.05 * a$; $R_2 = 0.37 * a$; $R_3 = 0.39 * a$; $R_4 = 0.45 * a$; $N_i = N_o = 8$; $\Delta = 0.1 * a$; $\gamma = 0.8$.**

**Section Three: Experimental Setup**

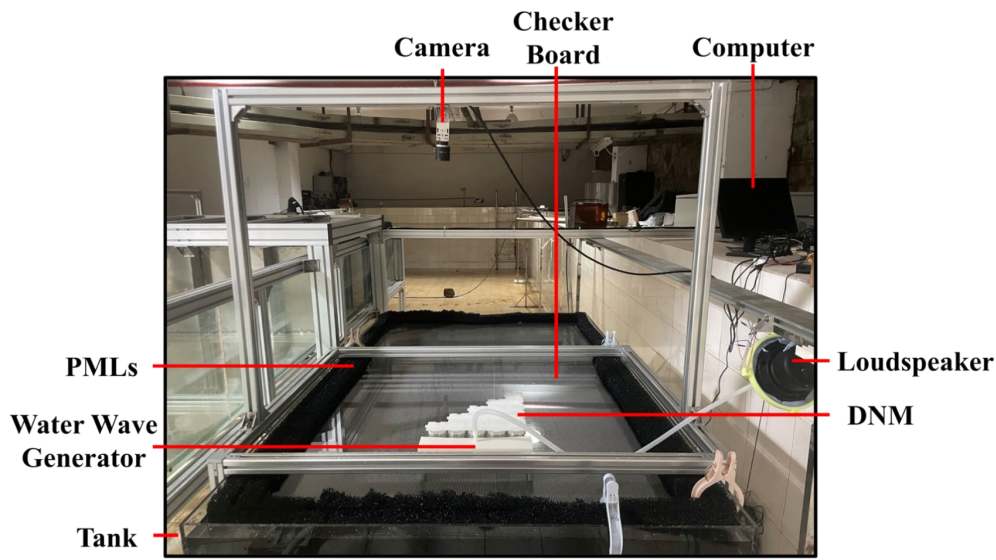

**Fig. S4 Experimental setup for fast checkerboard demodulation.**

In this experiment, Fast Fourier Demodulation (FCD) is applied to quantify and reconstruct the amplitudes of the water waves (*6*). The detailed set for experiment is shown in Fig. S4. Water waves are generated through a computer-controlled sinusoidal signal, allowing for the real-time control over both frequency and amplitude by computer. The vibrations made by the speaker travel through the air and are transferred to the wave generator. The energy is then converted from sound waves into water waves, which generate sinusoidal waves on the water surface.

To measure the wave amplitudes within the DNM, the FCD is used. The vibrations of the water waves cause disturbances in the static checkerboard pattern (checker board shown in as Fig. S4), originally used as a reference background. These distortions are compared with the reference image, and the differences are used to calculate the actual wave distribution. This process allows for the reconstruction of the displacement $\eta$ and velocity field $V$ of the water waves across the entire captured area.